# Aharonov-Bohm effect in graphene Fabry Pérot quantum Hall Interferometers

Yuval Ronen[1†], Thomas Werkmeister[2†], Danial Najafabadi[1], Andrew T. Pierce[1], Laurel E. Anderson[1], Young J. Shin[3], Si Young Lee[4], Young Hee Lee[4], Bobae Johnson[1], Kenji Watanabe[5], Takashi Taniguchi[6], Amir Yacoby[1,2], Philip Kim[1,2] *

**Quantum interferometers are powerful tools for probing the wave-nature and exchange statistics of indistinguishable particles. Of particular interest are interferometers formed by the chiral, one-dimensional (1D) edge channels of the quantum Hall effect (QHE) that guide electrons without dissipation. Using quantum point contacts (QPCs) as beamsplitters, these 1D channels can be split and recombined, enabling interference of charged particles. Such quantum Hall interferometers (QHIs) can be used for studying exchange statistics of anyonic quasiparticles. In this study we develop a robust QHI fabrication technique in van der Waals (vdW) materials and realize a graphene-based Fabry-Pérot (FP) QHI. By careful heterostructure design, we are able to measure pure Aharonov-Bohm (AB) interference effect in the integer QHE, a major technical challenge in finite size FP interferometers. We find that integer edge modes exhibit high visibility interference due to relatively large velocities and long phase coherence lengths. Our QHI with tunable QPCs presents a versatile platform for interferometer studies in vdW materials and enables future experiments in the fractional QHE.**

While interferometry techniques were originally accessible only in the domain of optics, in recent years electron-based interferometry has become a powerful probe of coherent quantum phenomena. One typical optical interferometry technique is using a FP interferometer to induce self-interference in a cavity formed between two reflectors.[1] In a 2-dimensional electronic system (2DES) in the QHE regime, QPCs that control the numbers of transmitting 1D electronic channels serve as tunable reflectors.[2] Combining two QPCs in the 2DES channel, FPs and related QHIs were realized in semiconductor heterojunctions.[3–11] In the FP QHI,[12] the magnetix flux contained in the area between the QPCs induces single-particle interference via the AB effect. It has been heavily investigated as a platform for topological quantum computation using the fractional QHE.[13] However, as both experimental and theoretical studies showed[9,14–16], Coulomb charging effects may obscure the AB interference signal, which has prevented realizations of the platform. While the charging effect is more significant in smaller FP interferometers, smaller FP interferometers are preferable, since shorter interference paths are more resilient against decoherence. This longstanding hurdle was addressed only in a recent study of FP devices where Coulomb interactions were suppressed by incorporating screening layers in proximity to the 2DES.[11] While this approach enabled observation of AB interference in the fractional QHE regime, the presence of global screening layers limits the versatility and tunability of the interferometers.

Graphene provides an alternative route of suppressing Coulomb interactions. Recent studies on ultraclean hBN/graphite encapsulated graphene vdW heterostructures report FQH states at moderate magnetic fields (< 10 T), as well as even-denominator fractional QHE states which may host non-Abelian anyons.[17,18] The large energy gaps of these states, ease of tuning the density, and, crucially, the ability to engineer the paths of the edge modes with local gates[19] make graphene vdW heterostructures a promising platform to realize versatile interferometers. Most importantly, the graphite gate and thin hBN dielectric layers serve to suppress charging effects without additional screening layers.

A few quantum-coherent devices have been previously fabricated in graphene-based vdW heterostructures.[20–26] While QHE intereference was observed in Mach-Zehnder interferometers (MZI) built across a graphene pn junction with co-propagating edge modes[20,25], random scattering at the physical edge served as uncontrolled beam splitters in these MZIs, limiting tunability and coherence. Furthermore, recent 'local probe' measurements have shown that edge modes in close proximity to etched graphene edges may suffer from dissipation due to counterflowing edge modes.[27] Therefore, to increase the edge mode coherence length, which is essential for high visibility interference, we electrostatically defined QHIs to enforce bulk separation from the physical edge of graphene. Additionally, the design allows an electrostatically defined sharp confining potential, which may prevent edge state reconstruction as well as maintain large velocity of the edge modes.[28]

[1]Department of Physics, Harvard University, Cambridge, MA 02138, USA  [2]John A. Paulson School of Engineering and Applied Sciences, Harvard University, Cambridge, MA 02138, USA  [3]Center for Functional Nanomaterials, Brookhaven National Laboratory, Upton, NY 11973, USA  [4]Center for Integrated Nanostructure Physics (CINAP), Institute for Basic Science (IBS), Suwon 16419, Republic of Korea.  [5]Research Center for Functional Materials, National Institute for Materials Science, 1-1 Namiki, Tsukuba 305-0044, Japan  [6]International Center for Materials Nanoarchitectonics, National Institute for Materials Science, 1-1 Namiki, Tsukuba 305-0044, Japan
[†]These authors contributed equally to this work *Corresponding author: pkim@physics.harvard.edu

Fig. 1a shows an electron microscope image of a representative device out of 4 devices we studied (see the method and supplementary information (SI) for fabrication details). A FP QHI requires two QPCs where edge channels are brought sufficiently close to induce backscattering, shown schematically in Fig.1b. There are 8 Ohmic contacts ($C_n$, n=1 to 8 as shown in Fig. 1b), four on each side of the FP interferometer, to source current and detect transmission and reflection by measuring the chemical potential of the QH edges. Each $i_{th}$ QPC ($i = 1,2$, respectively) may be described by two parameters: the probability of transmitting ($t_i$) and reflecting ($r_i = 1 - t_i$) a quasiparticle on the edge, where reflecting means backscattering to the opposite chiral edge. Neglecting phase-averaging and any decoherence processes, the probability for a quasiparticle emitted on the edge from C1 to transmit through both QPCs and reach the ground contact C4 (on the other side of the interferometer) is given by $T_{FP} = \frac{(1-r_1)(1-r_2)}{1+r_1 r_2 - 2\sqrt{r_1 r_2}\cos(\phi)}$, where $\phi$ is the phase acquired by a propagating particle in one revolution around the perimeter of the FP cavity.

Since graphene has no intrinsic bandgap, creating a QPC for edge states requires different implementation than in gapped semiconductors. For the ubiquitous QPC in a semiconductor, the Fermi level under the split gates needs to be set in the intrinsic band gap, depleting electronic states from the region. In graphene, we use the LL gaps that form in the QH regime in an analogous way. Partially-tunable transmission of QH edge modes in graphene was demonstrated using this operating principle.[29,30]

Fig. 1c shows transmission and reflection measurements for a QPC, demonstrating the operation of a single QPC device for $\nu_B = 2$ as we partition the inner edge (second LL) by adjusting gates in the range $1 < \nu_{QPC} < 2$. We measure the voltage difference $V_T$ ($V_R$) between the transmitted (reflected) edge states and the incoming edge states as shown in Fig. 1b. The transmission and reflection coefficients are related to the transmission resistance $R_T = V_T/I$ and reflection resistance $R_R = V_R/I$, where $I$ is total injected current. Current conservation guarantees $R_T + R_R = h/e^2 \nu_B$, and the reflection probability of the QPC is obtained from $r = (R_R \nu_B^{-1} e^2/h - \lfloor \nu_{QPC} \rfloor)$, where $\lfloor \nu_{QPC} \rfloor$ is the largest integer smaller than $\nu_{QPC}$. Each edge is partitioned separately by tuning the gates in the device simultaneously at a fixed magnetic field (8 T). Differnet QH edge modes can be transmitted through the QPC as we tune $\nu_B$ and accordingly $\nu_{QPC}$ (see S2 in SI). Even in $\nu_B = 6$ at a QPC width of 150 nm, we can see full transmission of all 6 edge modes (see Fig. S2), which means the average edge mode width is less than 12 nm. This mode width is comparable to the magnetic length, $\ell_B \equiv \sqrt{\hbar/eB} = 9$ nm for $B = 8$ T.

By cascading two QPCs in series, we construct a FP device (illustrated in Fig. 1b). In this FP device, each QPC is tuned using their respective split-gates and a common graphite bottom gate. In Fig. 1d, we first display line-cuts of $R_T$ for independently measured QPCs in the range of $0 < \nu_{QPC} < 2$ while the split gate voltages are set to $\nu_S = 0$. We observe the transmissions through the two QPCs are nearly identical in the plateau region but differ along the plateau transitions, where they show fluctuations. These transmission fluctuations differ from one thermal cycle to another, indicating that residual disorder configuration near the QPCs contributes to these variations. We observe fewer fluctuations in the single QPC device (Fig. 1c). This device has a thicker bottom hBN (74 nm) compared to the FP device (17 nm), and increased distance from the screening bottom gate reduces the likelihood for compressible states to form near the saddle point of the QPC potential, consistent with the proposed mechanism for transmission fluctuations as resonant charging of these compressible, localized states in the QPC region.[31]

Tuning the transmission through both QPCs simultaneously allows us to control the interference in the FPI. Our FP device uses a side *plunger gate* (with voltage $V_{PG}$) between the two QPCs to modify the QH edge mode trajectory within the FP cavity. Fig. 1e shows the simultaneously measured reflection and transmission across the FP as a function of $V_{PG}$. We set the bulk (including the FP cavity region) to $\nu_B = 2$ while $\nu_S = 0$ and partition the inner edge channel with QPC reflection probabilities $r_1 = 0.51$ (left) and $r_2 = 0.14$ (right). The outer edge channel passes fully through both QPCs. A clear oscillatory transmission $R_T$ and reflection $R_R$ are observed as a function of $V_{PG}$. They sum to a constant $h/e^2 \nu_B$, demonstrating that transport is governed solely by edge states. The normalized oscillation amplitude, visibility, is ~ 10 % at 30 mK and persists through 200 mK (inset of Fig. 1e), where visibility is defined as $(R_{max} - R_{min})/(R_{max} + R_{min})$.

The observed oscillatory behavior of $R_T(V_{PG})$ in the FP QHI can be attributed to the AB effect that modulates the interference phase $\phi$. As we scan $V_{PG}$, the enclosed effective area $A$ occupied by the interfering edge mode changes. At a fixed $B$, this area change $\delta A$ is related to the added (subtracted) charge $\delta Q = eB\delta A/\Phi_0$ in $\delta A$, where $\Phi_0 = h/e$. Since $\delta Q \approx C_{EG}\delta V_{PG}$, where $C_{EG}$ is the capacitance between the edge channel and plunger gate, the plunger gate modulates the total magnetic flux $\Phi = BA$ by $\delta\Phi = B\,\delta A$, leading to $\delta\phi = 2\pi\frac{\delta\Phi}{\Phi_0} \approx 2\pi C_{EG}\delta V_{PG}/e$.

An explicit demonstration of the AB interference in our QHI can be achieved by measuring the transmission/reflection through it as a function of variations in both the magnetic field, $\delta B$, and the plunger gate voltage, $\delta V_{PG}$. The expected phase evolution for a single revolution around the FP interferometer perimeter is given by $\delta\phi/2\pi \approx A\delta B/\Phi_0 + C_{EG}\delta V_{PG}/e$. Fig 2 shows the measured $R_T(B, V_{PG})$ in four operating regimes of the QPCs' reflection coefficients, from relatively open (Fig. 2a) to pinched (Fig. 2d). A periodically repeating stripe pattern (so-called pajama plot), whose constant phase (i.e., $\delta\phi = 0$) inclination agrees with 'AB interference', $\delta B/\delta V_{PG} = -\frac{C_{EG}\Phi_0}{eA} < 0$, is observed. The magnetic field periods, seen in the 2D-FFT (Fig. 2e-h), match an integer multiplicity of the enclosed flux in the lithographically defined area of $3\ \mu m^2$. The periodicity in $V_{PG}$ yields an edge-gate capacitance $C_{EG} = 16 \cdot 10^{-18}$ F. We correlate the visibility of the AB oscillations with the reflection coefficients of the QPCs in order to find a phase coherence length, assuming an exponential suppression (see SI3). For the inner edge of $\nu_B = 2$, this process yields a characteristic phase coherence length, $L_0 = 7.2\ \mu m$, on the order of the perimeter of the cavity, $L = 6.1\ \mu m$. We estimate electron temperature to be 60 mK. We remark that the AB oscillations observed in our FP QHI are robust. With various device sizes and designs, different $r_i$ values, and different filling fractions, we always observe a negative $\delta B/\delta V_{PG}$ slope in our pajama plots.

We note that even under the strongly pinched condition (Fig. 2d), the interference is 'AB dominated'. Measurements in similar area FP interferometers fabricated in GaAs structures often displayed a different behavior: lines of constant phase $\delta\phi = 0$ in the $B$ vs. $V_{PG}$ plane had zero (field independent) or positive slopes.[9,15] This complicated behavior of FP QHI, adopted the name 'Coulomb-dominated (CD)'[14], associated with strong Coulomb coupling between the interfering edge mode and the localized quasiparticle states in the bulk. The two regimes, AB and CD, can be understood employing a classical capacitive model. Defining $\xi \equiv C_{EB}/(C_{EB} + C_B)$, for the AB regime $\xi \to 0$, while for the CD regime $\xi \to 1$. Here, $C_{EB}$ is the capacitance between the interfering mode and the compressible puddle in the bulk, and $C_B$ is the total capacitance of the bulk puddle to ground.[14,16] Experimentally, there is a trade-off between making the interferometer smaller in order to increase particles coherence and minimizing the interaction parameter $\xi$. Our device, with two hBN insulating layers (with a relative dielectric constant $\epsilon = 4$), separating the bottom (top) graphite gate 17 (50) nm away, is estimated to have $\xi \ll 0.1$ and a charging energy scale $e^2/(2C_B) \approx 8\ \mu eV$, comparable to the state-of-art GaAs FP QHI where anyonic AB interference was reported.[11]

In our graphene FP QHI, the interfering edge is guided by a barrier set by the large LL gap underneath a biased graphite gate. This unique scheme allows us to investigate the decoherence mechanism of the edge modes. Fig. 3a shows a wide range of AB oscillations of the inner edge of $\nu_B = 2$ as a function of plunger gate voltage $V_{PG}$, where the LLs' filling fraction underneath the gate $\nu_{PG}$ varies between -2 and 3. The oscillations exhibit a reduced gate periodicity as the filling $\nu_{PG}$ increases. This a direct consequence of an increased $C_{EG}$ due to the edge mode moving closer to the plunger gate. The visibility of the AB oscillations does not change appreciably in the range $\nu_{PG} < 1$, suggesting that distance from the copropagating outer edge mode does not play a role in decohering the interfering inner edge mode. However, the visibility drops for $\nu_{PG} > 1$. Three different regimes appear in this range. First, for $1 < \nu_{PG} < 1.4$ (regime I) – the outer most edge mode moves away, while the interfering inner mode interferes with a slowly reduces visibility. For $1.4 < \nu_{PG} < 1.6$ (regime II), inner and outer modes are separated by a compressible region, which either decoheres the interfering mode or lowers the mode's velocity (due to a softer potential). For $1.6 < \nu_{PG}$ (regime III), the two edge modes approach the physical edge of the graphene layer, and the AB oscillations exhibits a visibility of more than an order of magnitude smaller. Indeed, the 2D Fourier transform of oscillating $R_T(B, V_{PG})$ (Fig. 3b) shows that the corresponding FP enclosed area increases from 3 µm² (regime I) to 7.5 µm² (regime III); equal to the combined area underneath the plunger gate and the FP cavity. We estimate the dephasing length $L_0 \sim 0.4$ µm for the portion of propagation along the etched graphene edge in regime III from the drop of the visibility (Fig. 3c). We attribute this strong dephasing to result due to the proximity of the interfering edge mode to the physical edge of the graphene layer. Indeed, a recent scanning probe study showed the presence of local counter propagating edge states at the QH regime as well as multiple dangling bonds at the physical edge.[27]

In order to enhance the visibility of FP QHI, one thus needs to engineer the edge states to increase $L_0$. In our graphene-based FP QHI, this goal can be achieved by (i) shielding the interference edge from the physical edge and charge puddles by guiding the edges via electrostatics and utilizing other QH edges for screening; and (ii) sharpening the electrostatic barrier potential to increase the edge mode velocity $v_{edge}$. The velocity of the interfering mode can be estimated via the ubiquitous 'lobe structure'.[32] Applying a finite source-drain bias $V_{SD}$ on the interfering mode produces additional modulation in $R_T(V_{SD})$, while all other parameters remain constant (Fig. 4a). Due to the self-interference condition, each time a full wave packet occupies the interferometer a constructive interference should occur yielding a phase shift of $2\pi = eV_{SD}L/\hbar v_{edge}$.[32,33,11] From the periodic modulation $R_T(V_{SD})$ at fixed $V_{PG}$ (upper panel in Fig. 4a), we estimate $v_{edge}= 7.6 \times 10^4$ m/s for the inner edge for $v_B = 2$ QH state. We further probe the interference on other edge modes, the extracted phase coherence lengths and velocities are summarized in Fig. 4b and in SI4. Interestingly, we find that screening of the interfering edge from both the etched physical edges and the bulk, by adding inner modes, say, in the configuration $(v_B, \lfloor v_{QPC} \rfloor) = (3, 2)$ improves both merits. We also tested a small range of the fractional regime. Fig. 5a shows a well developed $v_B = 7/3$ and $8/3$ (see SI5 for other FQH). The QPCs can partition integer QH edges as well as the inner fractional QH modes (Fig. 5b-c). While highly visible interference for the integer modes was observed (Fig. 5d), partitioning the fractional mode, did not result with visible AB interference.

In summary, our graphene-based FP QHI shows clear and robust AB dominated oscillations in the QHE regime. We extract a reduced coherence length for an edge mode propagating along the physical edge of graphene, demonstrating the importance of an electrostatically defined QHI. Experiments are ongoing to introduce independent control to the separated QPCs' gates, smaller perimeter devices, and lower electron temperature measurements to probe interference and coherence of fractional modes.

**Methods**
Graphene and hBN were mechanically cleaved from bulk crystal using thermal release tape. The tape containing exfoliated flakes was brought into contact with SiO$_2$ at 100℃ and baked for 1 minute. The tape then naturally cools to room temperature and was slowly removed from SiO$_2$ after 10 minutes. For the stacking procedure, a polycarbonate (PC) dry transfer method was used. PC film was made using 8 wt% solution; droplets of solution were squeezed between two glass slides and left to cure at room temperature. The transfer stamp was made by placing a small block of polydimethylsiloxane (PDMS) (Gel-pak) cut into a diamond shape (~$8 \times 5\ mm$) diagonal lengths on the glass slide and transferring the PC film on top of it (PC film extends laterally beyond than PDMS block, adhering to the glass slide). The stamp was baked at 180℃ for 20-30 minutes to ensure the film is pinned to the glass slide.

Stacking started with picking up a large flake of graphite (~$80 \times 80 \mu m$). The transfer stage was heated to 50℃ and graphite flake was pressed into contact with PC while lowering the stamp at a 1° tilt angle to the plane. The stage was then heated to 110℃ and cooled down to 80℃ with natural convection. During cool down the stamp was lifted mechanically to pick up the graphite flake. Subsequent layers were picked up by replicating the same procedure. It was crucial that subsequent layers were fully covered by previous layers to utilize the van der Waals force to assist in picking them up. After all the flakes (graphite-hBN-graphene-hBN-graphite) were picked up on the stamp with desired orientations, the stage was heated to 160-180℃ and the stack was laminated on SiO$_2$ in order to remove bubbles and hydrocarbons trapped in between the layers.[34] The stacking phase finished by placing the substrate containing the stack in Chloroform for a minimum of 3 hours, followed by annealing in vacuum at 300℃ to partially remove the polymer residue and enhance the adhesion to the substrate. The stack used for the FP device reported here had a top (bottom) hBN thickness of 50 (17) nm, while the single QPC device had a top (bottom) hBN thickness of 31 (74) nm.

Devices were fabricated using standard nanolithography processes, with stacks laminated on doped Si substrates with a 285-nm layer of SiO$_2$ that acted as a dielectric to the Si back gate that was used for contact doping. The device geometry was defined by reactive ion etching in O$_2$/CHF$_3$ using a polymethyl methacrylate (PMMA) resist (patterned by electron-beam lithography) as the etch mask. This etching was in two steps: first a pure O$_2$ etch of the top graphite, then a process with O$_2$/CHF$_3$ to etch through the entire stack where needed. Edge contacts to the exposed graphene were made by CHF$_3$ etching the hBN and thermal evaporation of 2/7/150 nm of Cr/Pd/Au at an angle with rotation.[35] Then, air bridges were patterned using a bilayer PMMA process followed by a short 20s O$_2$ plasma PMMA residue clean and thermal evaporation of 2/7/350 nm Cr/Pd/Au. For the FP interferometer device, the air bridge to contact the middle graphite gate region was not deposited at this step; it was deposited after the lines in the top graphite were etched. To etch the ~50 nm lines in the top graphite, a thinner PMMA resist was used and again a reactive ion etch with weak O$_2$ plasma alone was done in short ~30s steps. In between etches, the two-probe resistance between each bridge-contacted gate was checked until they were all separated, such that the hBN was minimally etched.

The 4 top graphite regions for a single QPC (Fig.1b insets) were separately controlled to set filling factors ν$_S$ and ν$_B$ under the split-gates and in the bulk regions on each side of the constriction, respectively, where LL filling factor $\nu \equiv n_e/n_\phi$, where $n_\phi = eB/h$ and $n_e$ is the areal electron density; $B$ is magnetic field perpendicular to the plane. At the region in the middle of the split-gates, where the graphite is etched away for a separation of ~150 nm, the electrostatics create a saddle-point potential at the QPC. This saddle point potential was carefully tuned using the bottom graphite gate (Section 1, SI). The number of edges transmitted through the saddle point, $\nu_{QPC}$, was most strongly controlled by the bottom gate, but it also exhibited a weak dependence on the top gate voltages as seen in each 2D plot of QPC operating points.

Experiments were performed in a Leiden wet dilution system with base temperature ~32 mK and estimated ~60 mK electron temperature. Unless otherwise noted, a constant 8 T perpendicular magnetic field was applied. Measurements were taken using standard lock-in amplifier techniques with an ac excitation current of 1 nA at 17.77 Hz applied to the sample. Bias dependence was taken by adding a DC current in series and afterwards integrated to give voltage.

**Acknowledgements**
We thank Bertrand Halperin, Moty Heiblum, Eli Zeldov, Hassan Shapourian, and Di Wei for helpful discussions. P.K., Y.R., T.W., and L.E.A. acknowledge support from DOE (DE-SC0012260) for measurement, characterization, and analysis. P.K., D.H., and Y.J.S. acknowledge support from DOE (DE-SC0019300) for sample preparation and characterization. K.W. and T.T. acknowledge support from the Elemental Strategy Initiative conducted by the MEXT, Japan, Grant Number JPMXP0112101001, JSPS KAKENHI Grant Number JP20H00354 and the CREST(JPMJCR15F3), JST. S.Y.L. and Y.H.L. acknowledge support from the Institute for Basic Science (IBS-R011-D1). T.W. and A.T.P. were supported by the Department of Defense (DoD) through the National Defense Science & Engineering Graduate (NDSEG) Fellowship Program. Nanofabrication was performed at the Center for Nanoscale Systems at Harvard, supported in part by an NSF NNIN award ECS-00335765. This research used resources of the Center for Functional Nanomaterials, which is a US DOE Office of Science Facility, at Brookhaven National Laboratory under contract number DE-SC0012704.

**Author contributions**
Y.R., T.W., and P. K. conceived the idea and designed the project. P.K. supervised the project. Y.R., T.W., and D.N. fabricated the devices. L.E.A., Y.J.S., B.J., S.Y.L., Y.H.L, and A.Y. helped and consulted in different stages of the fabrication process and analysis. Y.R., T.W., and A.T.P. performed the measurements. Y.R., T.W., A.Y., and P. K. wrote the paper with inputs from all authors.

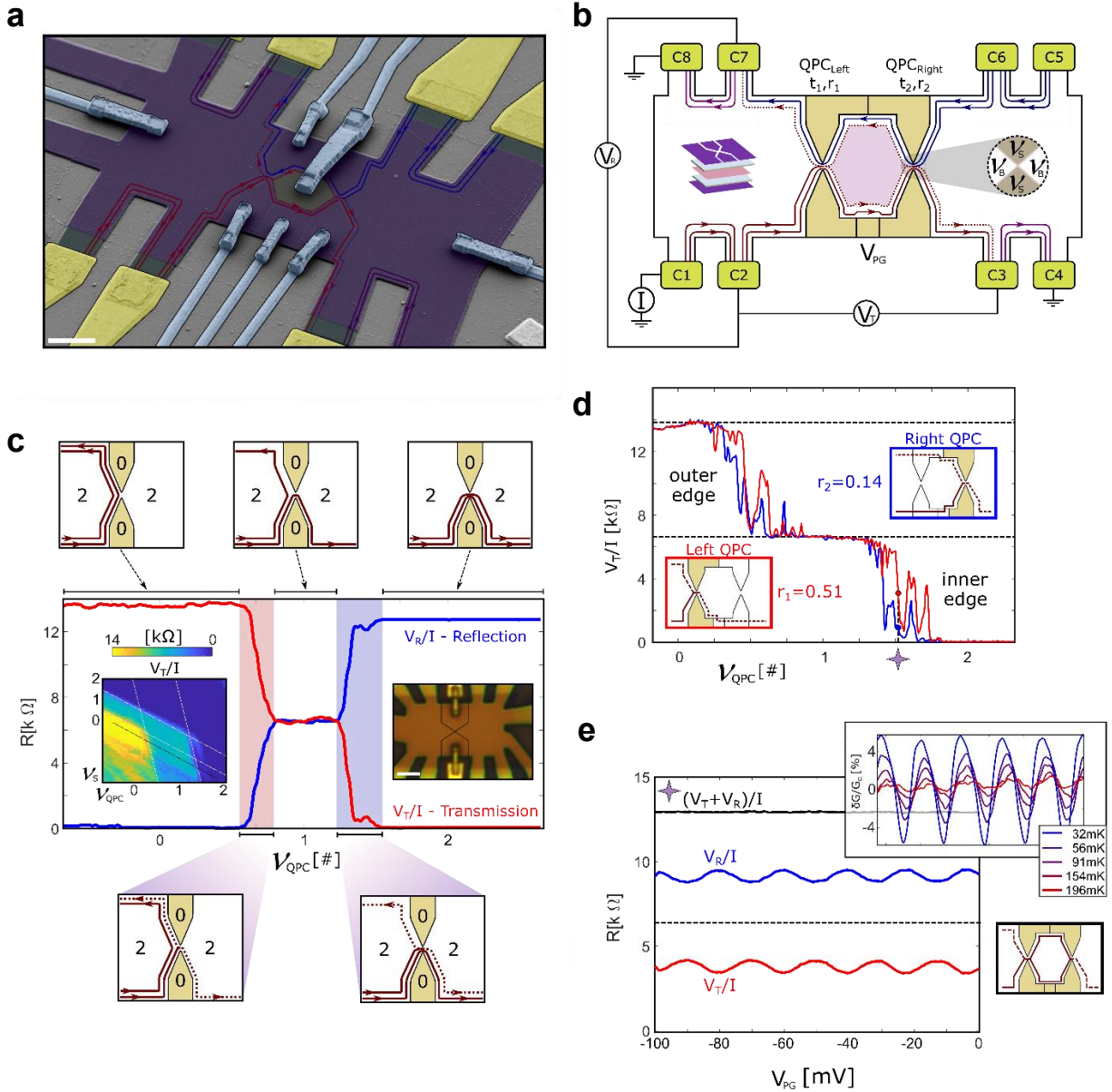

**Fig. 1 | Gate-defined Fabry-Pérot interferometer in graphene. a,** False color SEM image of a FP device. Contacts are yellow and bridges connecting to each region of the top graphite layer are blue. Scale bar: 2 $\mu m$. **b,** Schematic of a FP at filling factor 2 illustrating interference of the second LL edge (inner edge). Each QPC is realized by a pair of split gates, and a plunger gate (PG) tunes the area enclosed by the interfering edge (shaded red). For each QPC, the top graphite gates (inset illustrations) enable independent control of filling factors in the bulk $\nu_B$, split gates $\nu_S$, and QPC saddle points $\nu_{QPC}$. Current (1 nA) is injected into C1 while C4 and C8 are grounded. We measure $V_T = V_{23}$ and $V_R = V_{27}$. **c,** Fully tunable single QPC device (inset: optical image). Scale bar: 2 $\mu m$. A 2D map of QPC operating points for $\nu_B = 2$ is shown; $V_T/I$ as a function of $\nu_S$ and $\nu_{QPC}$, tuned by the top split gates and bottom graphite gate, respectively. The black solid line in the 2D map marks constant filling under the split gate, $\nu_S = 0$, and a continuous change in $\nu_{QPC}$. Line-cuts along the black line measuring $V_T/I$ (red) and $V_R/I$ (blue) demonstrate QPC operation for $\nu_B = 2$. **d,** Two adjacent QPCs showing overlap of edge partitioning regions in the $V_T/I$ measurement. **e,** FP interference at the inner edge of $\nu_B = 2$ as a function of PG voltage for QPCs operating point shown in fig. 1d. Inset: temperature dependence of the oscillation.

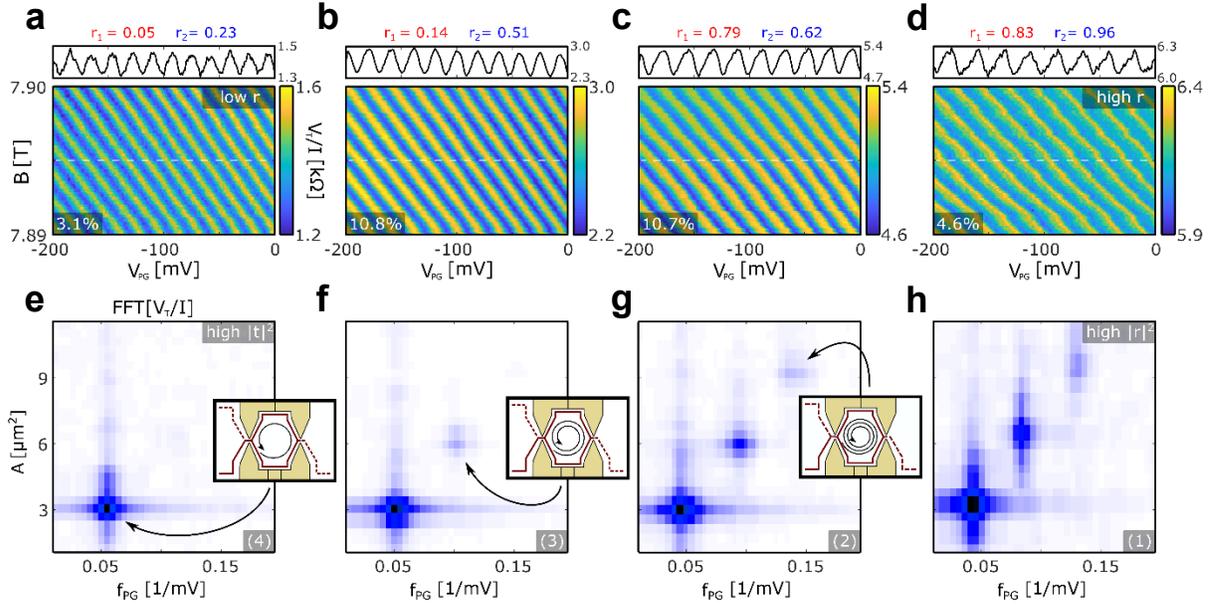

**Fig. 2 | Aharonov-Bohm (AB) dominated Fabry-Pérot interference. a-d,** Transmission resistance $R_T \equiv V_T/I$ oscillations as a function of magnetic field $B$ and plunger gate $V_{PG}$, showing clear AB oscillations. $\nu_B = 2$ and $\nu_S = 0$ and we observe the interference of the inner edge state. Reflection values of each working point of the QPCs are stated above each figure and visibility in percentage at the lower-left corner. In the upper panel of each plot we show a line cut as a function of the plunger gate along the white dashed line. **e-h,** 2D-fast Fourier transform (FFT) of the transmission resistance plots A-D, respectively, as a function of area and plunger gate periodicity. In the insets of figs. F, G, and H we illustrate the origin of each peak in the FFT signal. Higher harmonics appear when the QPCs are pinched (large $r_i$), physically corresponding to contributions from single-particle trajectories that make multiple revolutions around the interferometer area.

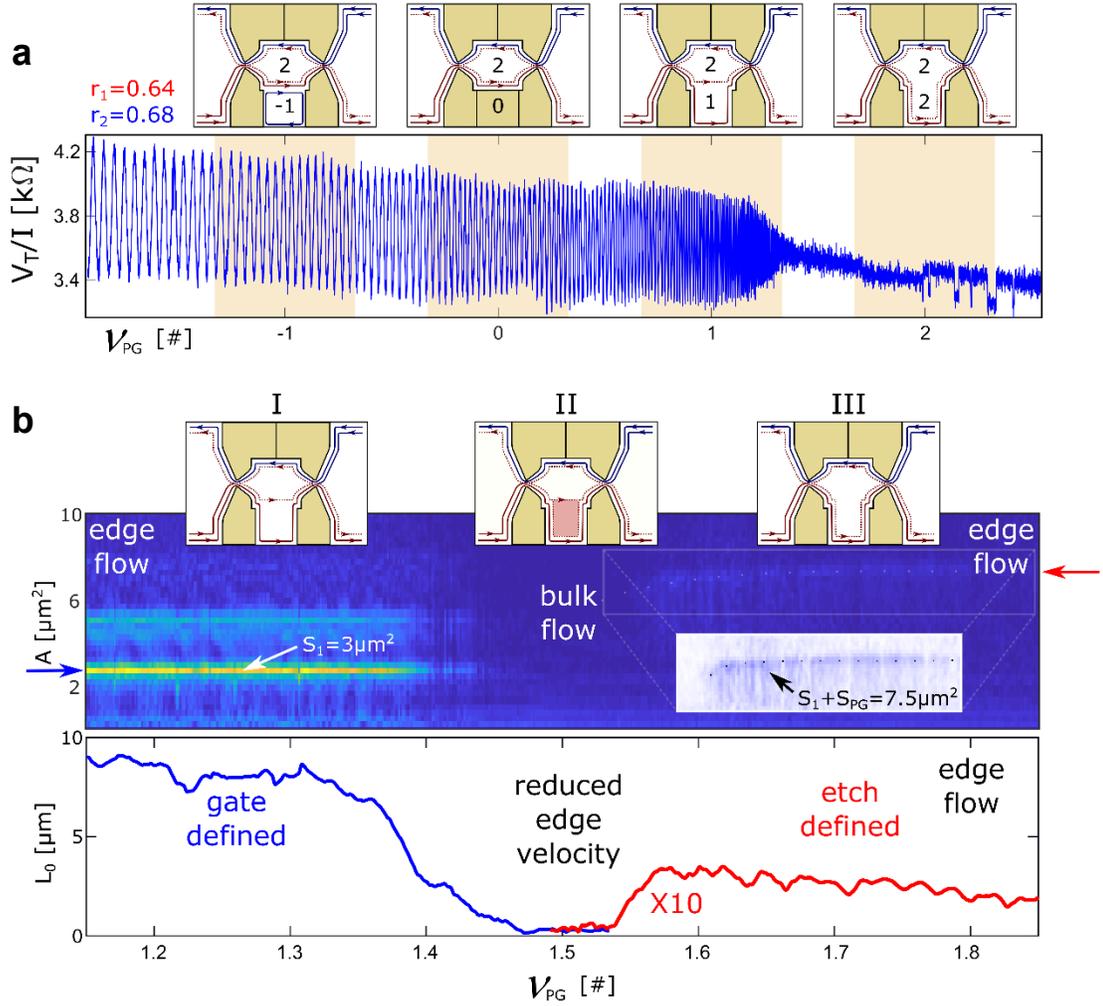

**Fig. 3 | Gate vs. etch defined interferometer: a,** Transmission resistance $R_T$ oscillations as a function of the filling factor under the plunger gate $\nu_{PG}$ spanning edge propagation along filling factor -1 to 2. As $\nu_{PG}$ increases, the interference edge capacitance to the plunger gate increases, thereby reducing the oscillation period. As $\nu_{PG}$ transitions from 1 to 2, visibility drops considerably owing to propagation of the edge along the etched graphene. **b,** 2D FFT of $R_T$ as a function of area and plunger gate filling showing 3 distinct regions: I. Interference is gate-defined, showing the expected area of $3\ \mu m^2$ (blue arrow) and additional harmonics. II. Suppressed oscillation due to bulk conductance under the plunger gate. III. Suppressed interference region due to etch-defined propagation of the interfering edge matching a fabricated area of $7.5\ \mu m^2$ (red arrow). **c,** Extracted coherence length as a function of plunger gate filling for regions I (blue) and III (red) for $3\ \mu m^2$ and $7.5\ \mu m^2$ areas, respectively.

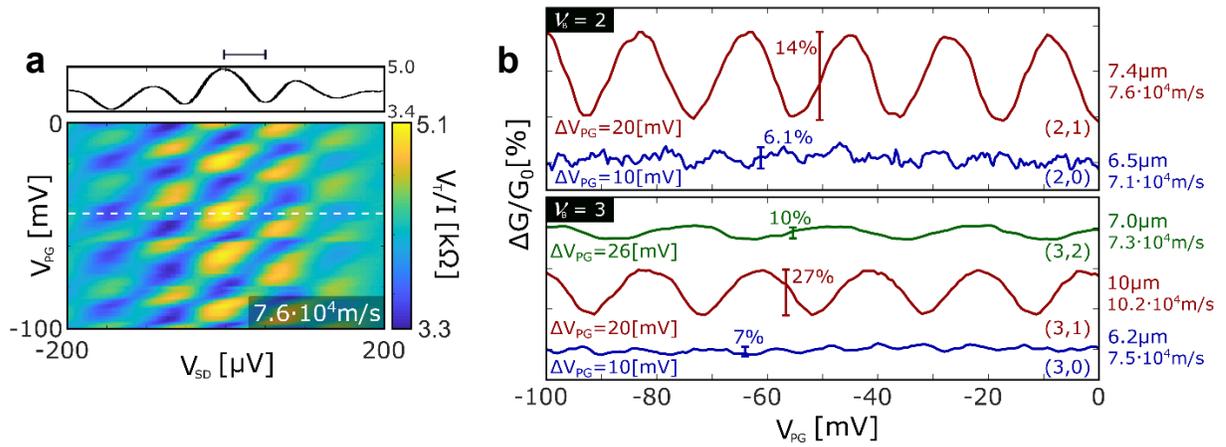

**Fig. 4 | Edge mode velocity and comparison of oscillations in different filling factors. a,** Transmission resistance $R_T$ oscillations as a function of $V_{SD}$ and $V_{PG}$ showing a checkerboard pattern. Edge mode velocity is estimated from the lobe structure. Upper panel show a cut of the data along the white dashed line. **b,** Comparing oscillation in the different edges at filling factor 2 (upper) and 3 (lower) as a function of plunger gate. Plunger gate periodicity, visibility, edge mode velocity and coherence length are written next to each plot. For more information on PG periods, see SI4.

x

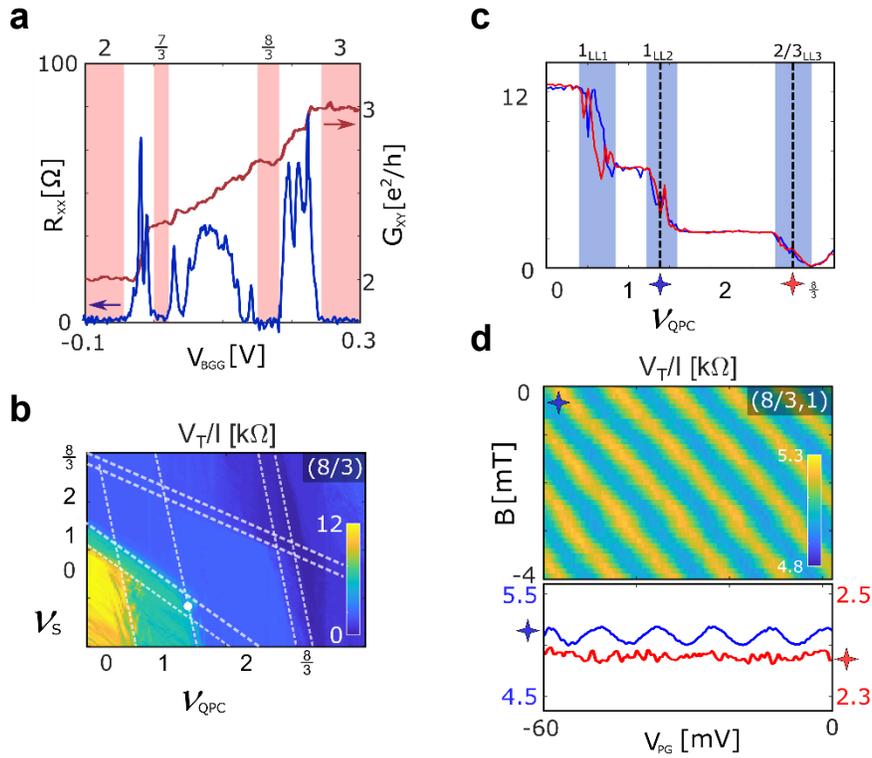

**Fig. 5 | Aharonov-Bohm interference of an integer edge when the bulk is in a fractional filling. a,** Measurement of $R_{XX}$ and $R_{XY}$ demonstrating fully developed FQH states at $\nu_B = \frac{8}{3}$ and $\frac{7}{3}$. **b,** 2D map of the operating points of the QPC at $\nu_B = \frac{8}{3}$; $V_T/I$ as a function of $\nu_S$ and $\nu_{QPC}$, tuned by the top split gates and bottom graphite gate, respectively. **c,** $V_T/I$ of the left and right QPC showing integer and fractional edge partitioning. **d,** Tuning to the marked point in (c), we measure interference of the innermost integer edge. Lower panel: $V_T/I$ measurement to an integer (blue) and a fractional (red) edge for $\nu_B = \frac{8}{3}$.

# Supplementry information:

# Aharonov-Bohm Effect in Graphene Fabry–Pérot Quantum Hall Interferometers

Yuval Ronen[1†], Thomas Werkmeister[2†], Danial Najafabadi[1], Andrew T. Pierce[1], Laurel E. Anderson[1], Young J. Shin[3], Si Young Lee[4], Young Hee Lee[4], Bobae Johnson[1], Kenji Watanabe[5], Takashi Taniguchi[6], Amir Yacoby[1,2], Philip Kim[1,2] *

[1] Department of Physics, Harvard University, Cambridge, MA 02138, USA

[2] John A. Paulson School of Engineering and Applied Sciences, Harvard University, Cambridge, MA 02138, USA

[3] Center for Functional Nanomaterials, Brookhaven National Laboratory, Upton, NY 11973, USA

[4] Center for Integrated Nanostructure Physics (CINAP), Institute for Basic Science (IBS), Suwon 16419, Republic of Korea.

[5] Research Center for Functional Materials, National Institute for Materials Science, 1-1 Namiki, Tsukuba 305-0044, Japan

[6] International Center for Materials Nanoarchitectonics, National Institute for Materials Science, 1-1 Namiki, Tsukuba 305-0044, Japan

[†] These authors contributed equally to this work

* Corresponding author: pkim@physics.harvard.edu

# SUPPLEMENTARY SECTION 1: FABRICATION PROCESS

The PC stamp is made of a glass slide, PDMS (gel-pak), and PC film as shown in Supp. Fig. 1-1 (A). In this method the PDMS is cut into a diamond shape $\sim 40 mm^2$, placed on the corner of glass slide and covered by PC film. The stamp is baked at $180°C$ for 20-30 minutes. After baking, the adhesion between PC-Glass slide is higher than adhesion between PC-PDMS, therefore selecting a larger PC film than PDMS will ensure the film remains fully attached during transfer. The main advantage of a diamond shaped stamp compared to standard square stamp is that is diamond stamp has 50% less contact area compared square stamp. This reduced contact area decreases the probability of pc film getting stuck on the substrate and failure of the transfer.

The general transfer method for each layer is shown in Supp. Fig. 1-1 (B). For each pick up the contact is initiated at $50°C$ and a $1°$ tilt angle. The flake is brought into contact with the stamp while the temperature is raised to $\sim 100 - 110°C$, and the flake is picked up during natural convection cool down at a temperature range bellow $90°C$. In our transfer method, graphite is used for the top layer. Compared to the more common hBN assisted pick up method (i.e. using hBN as the top layer), this simplifies the etching process when fabricating our devices. Optical images of the stack assembly progress are shown in supp. Fig. 1-1 (C-H). Selecting a large flake of graphite as the top layer provides an assistive van der Waal force for picking up subsequent layers.

One of the most critical steps in this method is the drop sequence. After all the layers are picked up on the stamp, the stage is heated to $\sim 160°C$ - $180°C$ and the stack is laminated to the substrate as shown in supp. Fig. 1-2. This method has proven to remove the bubbles with an approximately 70% success rate. However, it requires precise control of wave front to achieve a sufficiently slow and steady rate of expansion, which remains challenging.

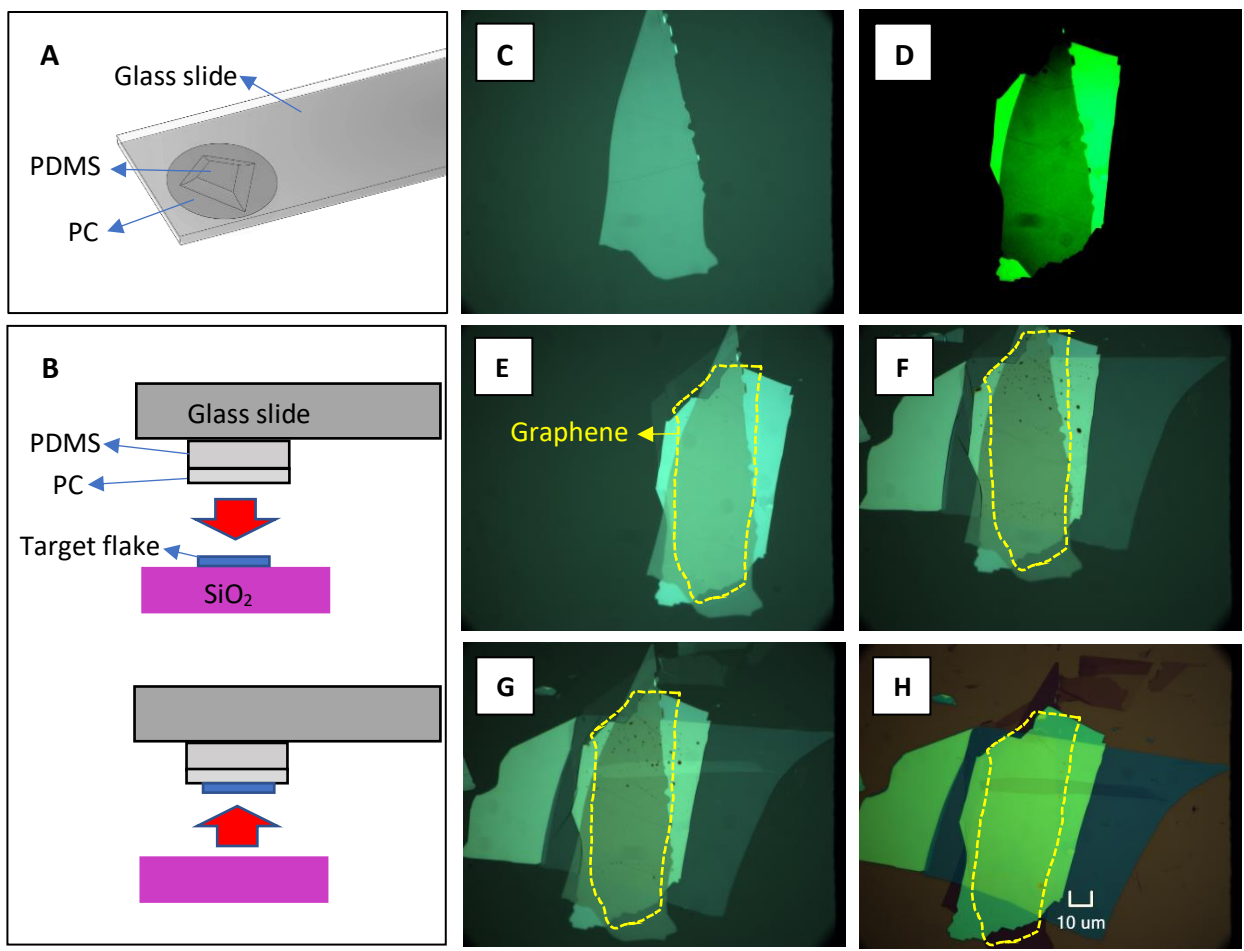

**SUPP. FIG. 1-1. PC Transfer stamp and stacking sequence. (A)** Components of stamp. **(B)** Pick up process. The stamp is tilted ~1° alongside longitudinal axis of diamond shape contact area (not shown in figure). **(C)** Optical image of graphite on the stamp after picking it up. It is common to see folds or cracks on graphite due to the large lateral area of the flake and thermal expansion, but they are not fixed until the last step. **(D)** Optical image of graphite-hBN on stamp. **(E)** Graphite-hBN-graphene on stamp. **(F)** Graphite-hBN-graphene-hBN on stamp. Most bubbles are formed during this step at the interface of graphene. **(G)** Optical image of stack graphite-hBN-graphene-hBN-graphite on stamp. **(H)** Stack on SiO$_2$ after laminating at 160℃ to 180℃. During this step bubbles are pushed to the edges of the flakes or accumulate at defects and folds.

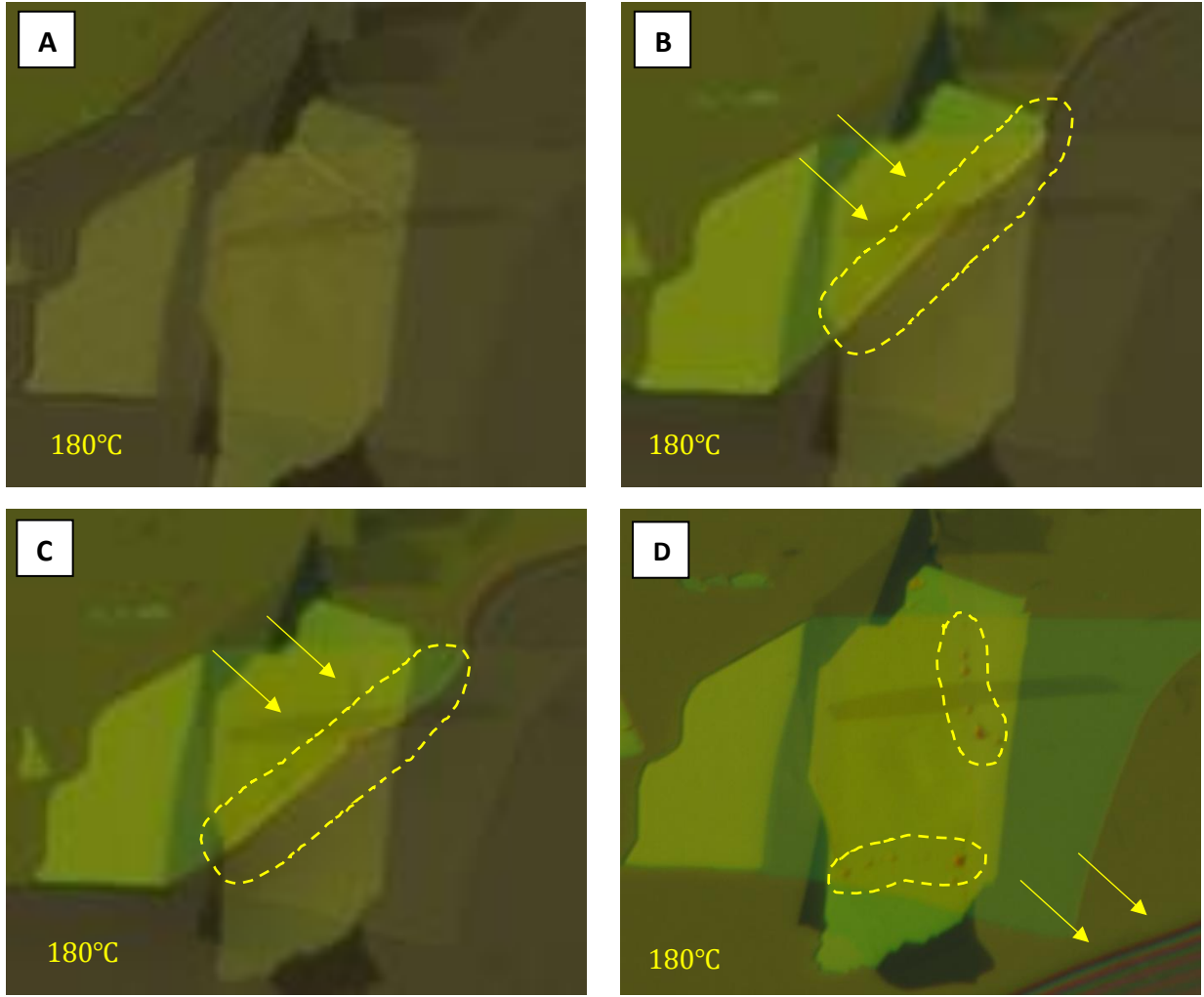

**SUPP. FIG. 1-2. Snapshots of stack during lamination on SiO₂.** The stamp is tilted ∼1° and mechanically pressed onto a SiO₂ surface at fixed temperature of 180℃. Arrows indicate the direction of wave front movement. **(A)** The wave front of the PC in contact with the SiO2 contact approaches the stack. **(B-C)** As the wave front smooths out at temperatures above the glass transition temperature of PC, the visible bubbles and fold defects are pushed along the contact front. **(D)** When the stack has been fully pressed onto the substrate, bubbles have been moved and pinned to the outer edge of graphene flake. A second line of bubbles are visible at the bottom of the stack where the graphene was cracked.

After removing the PC residue in Chloroform and annealing the stack at > 300°C for 3h to ensure that it adheres to the substrate and will remain mechanically and chemically stable through subsequent processing, the nanolithography processes outlined in methods are followed to fabricate a device. Supp. Fig. 1-3 shows the fabrication process of a single QPC in flowchart form.

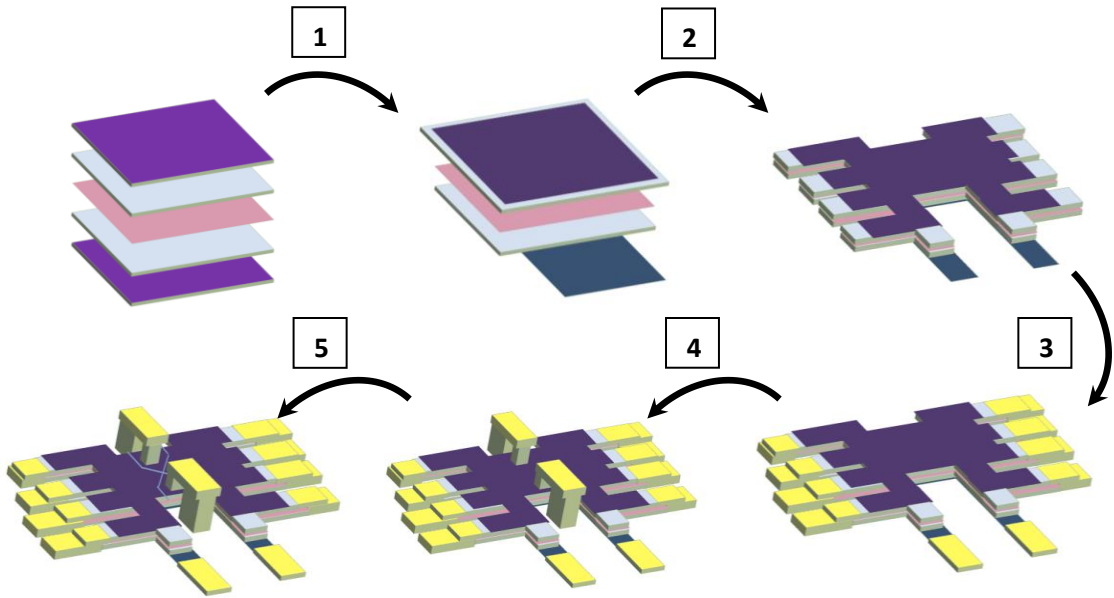

**SUPP. FIG. 1-3. Fabrication process schematic for QPC. (1)** Etch the top graphite into the desired shape; it must extend from the bottom graphite (recolored blue here) to avoid drastic filling factor changes or PN junctions from forming at the contacts, since we use the Si back gate to dope electrons and the Cr/Pd/Au edge contacts naturally dope electrons. **(2)** Etch through the entire stack to define desired geometry and distinct regions for contacts. **(3)** Deposit edge contacts to the exposed graphene and bottom graphite regions, as well as leads to the bridge locations. **(4)** Deposit gold air bridge contacts to top graphite. Note: this device would also have 2 additional bridges to contact the other regions that are separated after the next step. **(5)** Etch ~50nm lines into the top graphite to define the split-gates, using the process described in methods.

## SUPPLEMENTARY SECTION 2: QPC OPERATION

The main text (Fig. 1(B)) demonstrates a single QPC device operating with bulk filling factor $\nu_B = 2$. In Supp. Fig. 2, we show a similar QPC operating in $\nu_B = 1$ (A, D); $\nu_B = 2$ (B, E); and $\nu_B = 6$ (C, F). Thus, we demonstrate control of the QPC transmission over a wide range of LL's edge mode configurations that may be of interest.

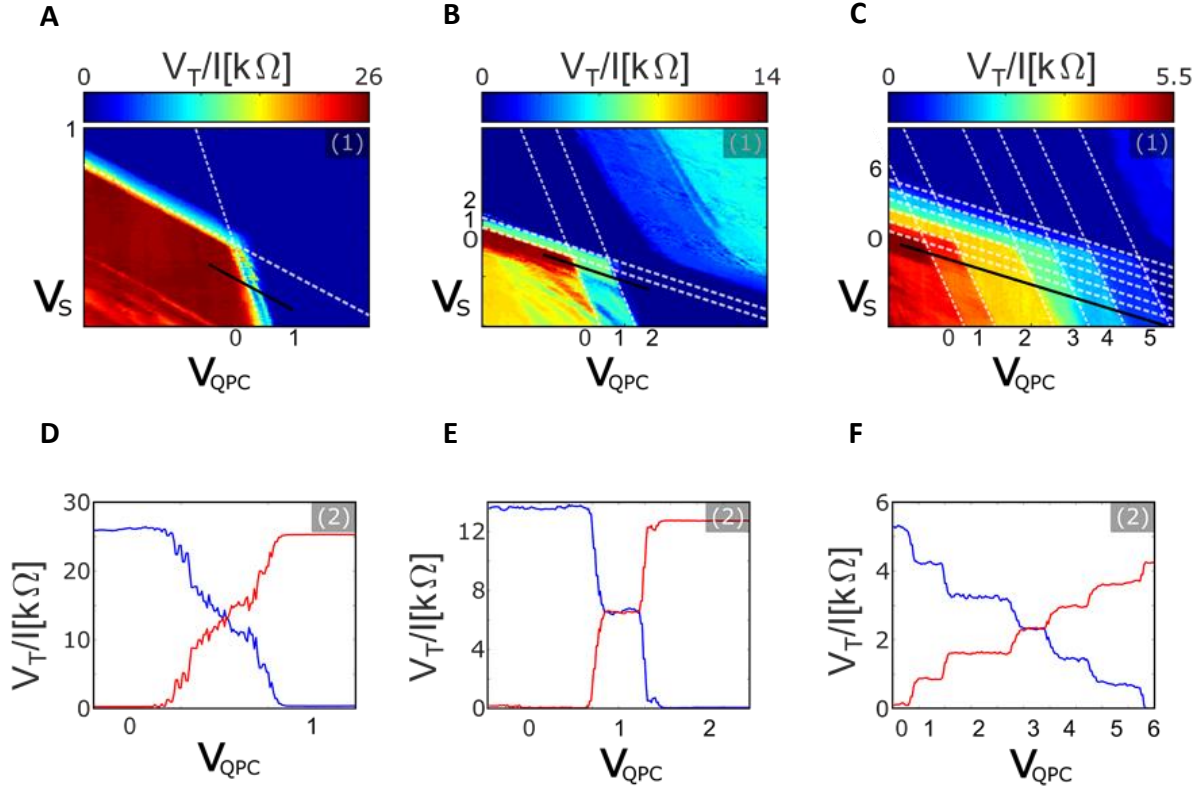

**SUPP. FIG. 2. QPC operating at various bulk fillings.** 2D maps of the phase-space of the QPC for **(A)** $\nu_B = 1$; **(B)** $\nu_B = 2$; and **(C)** $\nu_B = 6$ are shown; $V_T/I$ as a function of $\nu_S$ and $\nu_{QPC}$, tuned by the top split gates and bottom graphite gate, respectively. The black-solid line in the 2D map represents a constant filling under the split gate, $\nu_S = 0$, and a continuous change in $\nu_{QPC}$. Line-cuts along the black line measuring $V_T/I$ and $V_R/I$, blue and red, respectively, are shown for **(D)** $\nu_B = 1$; **(E)** $\nu_B = 2$; and **(F)** $\nu_B = 6$.

## SUPPLEMENTARY SECTION 3: FINITE PHASE COHERENCE

By including a finite phase coherence length $L$ (precisely a length for dephasing by a factor $e^{-1}$) into the single-particle model, we find that the total transmission probability from the Fabry-Pérot (FP) interferometer is given by

$$T_{FP} = \frac{(1-R_1)(1-R_2)}{1+R_1 R_2 e^{\frac{-2P}{L}} - 2\sqrt{R_1 R_2}\cos(\theta) e^{\frac{-P}{L}}} \tag{S1}$$

where $P$ is the perimeter of the interference loop. $P = 6.1\ \mu m$ for the FP device demonstrated in the paper and in the simulations shown. As a function of $R_1$ and $R_2$, the visibility of oscillations is then proportional to

$$V(R_1, R_2; L) = \frac{(1-R_1)(1-R_2)}{1+R_1 R_2 e^{\frac{-2P}{L}} - 2\sqrt{R_1 R_2} e^{\frac{-P}{L}}} - \frac{(1-R_1)(1-R_2)}{1+R_1 R_2 e^{\frac{-2P}{L}} + 2\sqrt{R_1 R_2} e^{\frac{-P}{L}}} \tag{S2}$$

which we plot as a percentage of $Max\{V(R_1, R_2; L \to \infty)\} = 1$ for various values of the phase coherence length $L$ in Supp. Fig. 3-1. We see generally that the maximum visibility is achieved for $R_1 = R_2 \equiv R$ and that in the case of infinite phase coherence ($L \to \infty$) the oscillations achieve maximum visibility as $R \to 1$. Moreover, as $R \to 1$ the shape of oscillations becomes sharp, as interference terms corresponding to multiple revolutions around the loop all contribute fully to the coherent sum. However, as $L$ is reduced to smaller finite values, the most visible configuration shifts to $R = R_{MAX} < 1$. As $L$ is reduced, the contributions of multiple revolution paths to the interference are suppressed exponentially, which modifies the visibility for all values of $R_1$ and $R_2$. By fitting the amplitude of oscillations normalized by the expected oscillation magnitude (i.e. the measured visibility) to the theoretical visibility as a function of $R_1$ and $R_2$, we may extract characteristic phase coherence length $L$. We show the fit for the inner edge of $\nu_B = 2$ in Supp. Fig. 3-2, from which we extract $L = 8.1\ \mu m$.

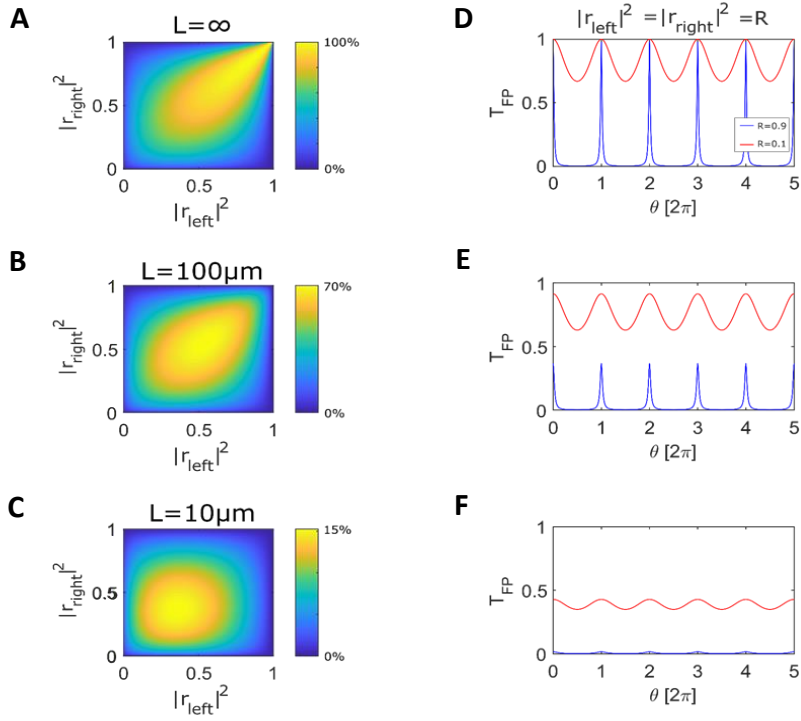

**SUPP. FIG. 3-1. Theoretical plots of visibility and oscillation shape with finite phase coherence.** The visibility $V(R_1, R_2; L)$ from eq. (S2) is plotted as a percentage of $Max\{V(R_1, R_2; L \to \infty)\} = 1$ versus $R_1 = |r_{\text{left}}|^2$ and $R_2 = |r_{\text{right}}|^2$ for **(A)** $L \to \infty$, **(B)** $L = 100\mu m$, and **(C)** $L = 10\mu m$. In all plots we have $P = 8.1\mu m$ fixed, which is the actual perimeter of our FP interferometer. $R_{MAX}$ and the maximum visibility $V(R_{MAX}, R_{MAX}; L)$ both shift to smaller values as $L$ is reduced, and the decay of the visibility away from the maximum point is uniquely determined by $L$. For two characteristic values $R = 0.9$ (blue) and $R = 0.1$ (red), we plot the shape of the oscillations with $\theta$ in eq. (S1) for the corresponding situations **(D)** $L \to \infty$, **(E)** $L = 100\mu m$, and **(F)** $L = 10\mu m$. $R \to 1$, the oscillations are sharp due to contributions from multiple revolutions, but the amplitude is exponentially suppressed with decreasing $L$. $R \ll 1$, the oscillations are sinusoidal, eventually the contribution from 1 revolution effectively contributes.

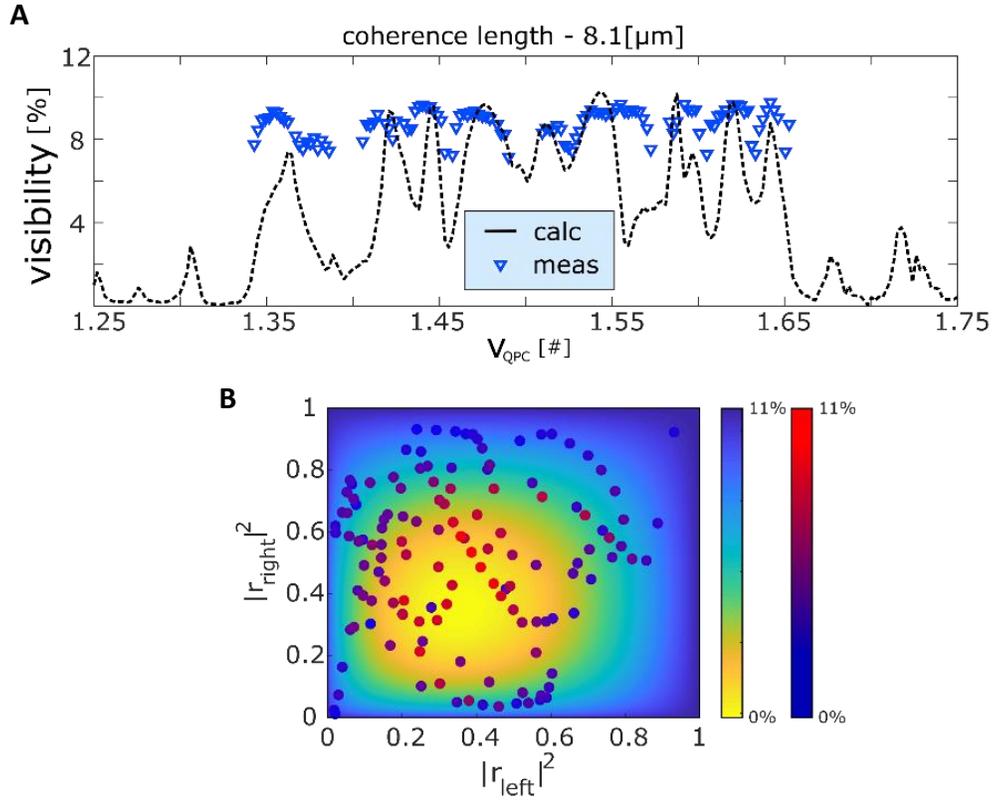

**SUPP. FIG. 3-2. Extracting phase coherence length from visibility. (A)** With $\nu_B = 2$, the inner edge (second LL) is partitioned by setting $1 < \nu_{QPC} < 2$. Here we show the visibility of the measured oscillations as a percentage of the visibility in the perfect phase coherence case (i.e. percentage of $6k\Omega$), plotted versus $\nu_{QPC}$ of one of the QPCs. The other QPC is also varying similarly, though not exactly the same, along this scan. Using the extracted characteristic phase coherence length $L = 8.1\mu m$ we calculate expected visibility, which fits reasonably to the measured points, particularly at intermediate transmissions. **(B)** We show the theoretical plot for $L = 8.1\mu m$ with a scatterplot of each data point that we measured as a function of $R_1 = |r_{left}|^2$ and $R_2 = |r_{right}|^2$. The scatter points fit best to this theoretical plot. Hence, we extract $L = 8.1\mu m$, and repeating this fit method for different edge allows comparison between coherence lengths.

**SUPPLEMENTARY SECTION 4: INTERFERENCE IN VARIOUS BULK FILLINGS**

We observe Aharonov-Bohm interference for all integer edge modes in $\nu_B = 2$ and $\nu_B = 3$, as summarized in Supp, Fig. 4-1. The magnetic field values shown are relative to 8T, and $V_{PG}$ is relative to a filling $\nu_{PG} = 0$ under the plunger gate.

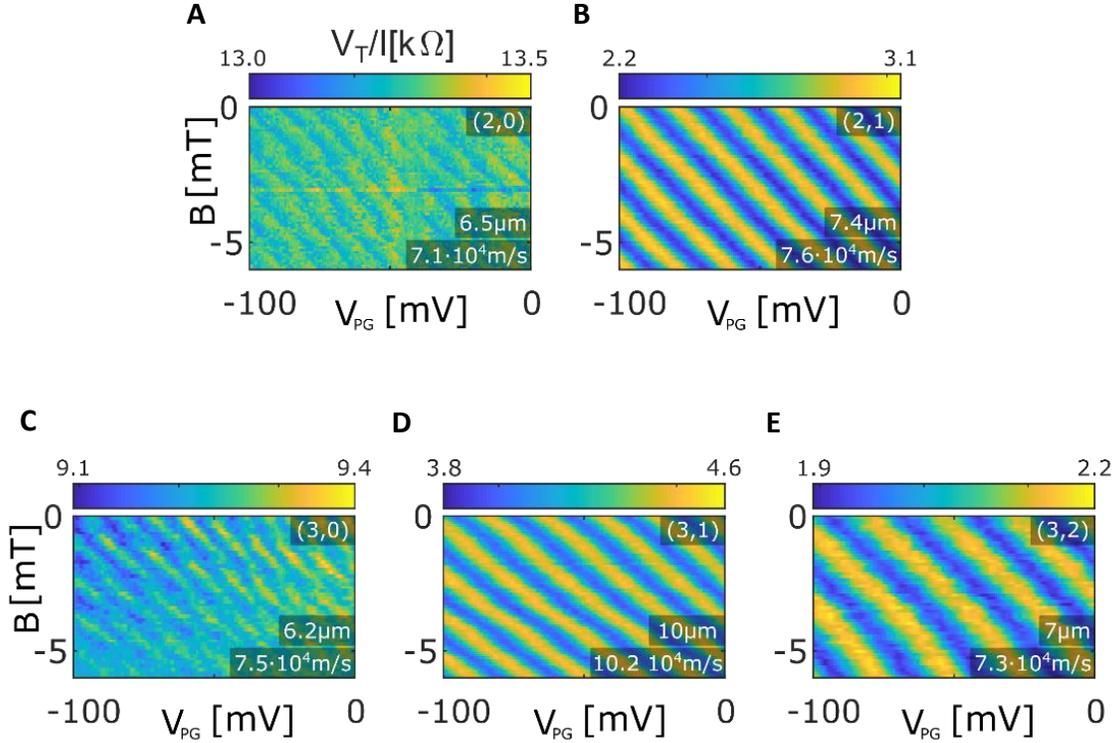

**SUPP. FIG. 4-1. Aharonov-Bohm interference in the B field, $V_{PG}$ plane for various edges. (A)** Outer edge of $\nu_B = 2$; $0 < \nu_{QPC} < 1$. **(B)** Inner edge of $\nu_B = 2$; $1 < \nu_{QPC} < 2$. **(C)** Outer edge of $\nu_B = 3$; $0 < \nu_{QPC} < 1$. **(D)** Middle edge of $\nu_B = 3$; $1 < \nu_{QPC} < 2$. **(E)** Innermost edge of $\nu_B = 3$; $2 < \nu_{QPC} < 3$. Importantly, we see that the maximum characteristic coherence length and edge mode velocity (both inset in bottom-right) is achieved for the middle edge of $\nu_B = 3$, where the interfering edge is screened by adjacent edges from decohering interactions in both the bulk and at the gate-defined edge. The area of the interferometer shrinks by nearly a factor of 2 from the outermost edge to the innermost, as seen from the magnetic field period approximately doubling.

According to eq. (2) of the main text, the plunger gate period is inversely proportional to the mutual capacitance of the interfering edge and the plunger gate $C_{eg}$. In Fig. 3(A) we saw that the oscillation period decreases as $\nu_{PG}$ increases, since $C_{eg}$ increases as the edge channel moves closer to the PG. Moreover, in Fig. 4(B) the outermost edge, closest to PG, shows the smallest period, corresponding to the largest $C_{eg}$, while the higher LL edges show progressively larger periods, corresponding to smaller $C_{eg}$. The PG periods for different edges are summarized in Supp. Fig. 4-2. We attribute the large difference in $C_{eg}$ to both different spatial separations of the interfering edge to the gate as well as screening of this capacitive coupling by adjacent edges.

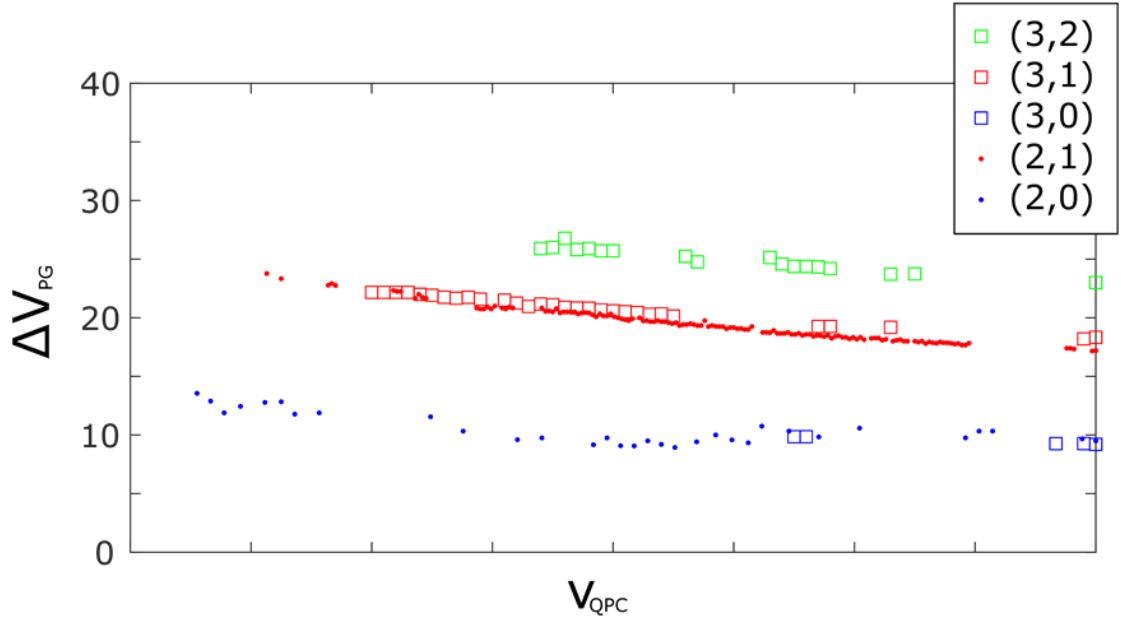

**SUPP. FIG. 4-2. Plunger gate periodicity $\Delta V_{PG}$ as a function of $\nu_{QPC}$ for various edge configurations ($\nu_B$, n).** The horizontal axis spans from $n < \nu_{QPC} < n + 1$ for each configuration, i.e. for (3,1) the middle edge mode of $\nu_B = 3$ is partitioned and $1 < \nu_{QPC} < 2$, where both QPCs roughly moving in coordination. The slow decrease as $\nu_{QPC}$ increases arises from an overall increase in $C_{eg}$ as the PG scans to more negative voltage to maintain $\nu_{PG} = 0$, sharpening the electrostatic boundary between the PG region and bulk.

**SUPPLEMENTARY SECTION 5: INTERFERENCE AT FRACTIONAL FILLING**

In addition to Fig. 5(A-D) of the main text, we also observe Aharanov-Bohm interference of the nearest integer edge mode when the bulk is in $\nu_B = \frac{10}{3}$, summarized in Supp. Fig. 5.

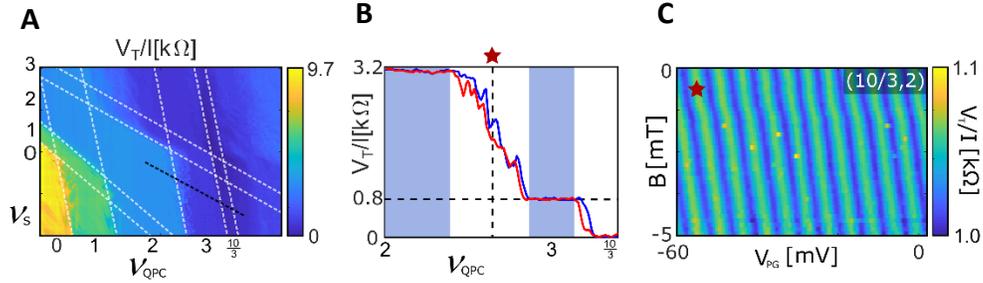

**SUPP. FIG. 5. Integer interference with fractional bulk filling:** **(A)** QPC phase-space in bulk filling factor $\nu_B = {}^{10}/_3$ as a function of $\nu_S$ and $\nu_{QPC}$. Black dashed line demonstrates QPC operation where the split gates are at filling factor $\nu_S = 2$. **(B)** $R_T$ of the left and right QPC along the black dashed line in fig 4d, showing the expected values of the integer and fractional edge portioning. **(C)** Aharonov-Bohm dominated resistance oscillations in the third integer LL in bulk filling factor $\nu_B = {}^{10}/_3$. Working point of the QPCs is depicted by the red star in (B).